\documentclass[journal]{vgtc}                     

\onlineid{0}

\vgtccategory{Research}

\vgtcpapertype{please specify}

\title{Roses Have Thorns: Understanding the Downside of Oncological Care Delivery Through Visual Analytics and Sequential Rule Mining
}

\author{%
 Carla Floricel, Andrew Wentzel, Abdallah Mohamed, C.David Fuller, Guadalupe Canahuate, and G.Elisabeta Marai
}

\authorfooter{
  \item
  	C. Floricel, A. Wentzel, and G.E. Marai are with the University of Illinois Chicago. E-mail: {cflori3@uic.edu | gmarai@uic.edu}.
  \item
  	G. Canahuate is with the University of Iowa. 

  \item A. Mohamed and C.D. Fuller are with the M.D. Anderson
Cancer Center at the University of Texas.
}

\abstract{%
  Personalized head and neck cancer therapeutics have greatly improved survival rates for patients, but are often leading to understudied long-lasting symptoms which affect quality of life. Sequential rule mining (SRM) is a promising unsupervised machine learning method for predicting longitudinal patterns in temporal data which, however, can output many repetitive patterns that are difficult to interpret without the assistance of visual analytics. We present a data-driven, human-machine analysis visual system developed in collaboration with SRM model builders in cancer symptom research, which facilitates mechanistic knowledge discovery in large scale, multivariate cohort symptom data. Our system supports multivariate predictive modeling of post-treatment symptoms based on during-treatment symptoms. It supports this goal through an SRM, clustering, and aggregation back end, and a custom front end to help develop and tune the predictive models. The system also explains the resulting predictions in the context of therapeutic decisions typical in personalized care delivery. We evaluate the resulting models and system with an interdisciplinary group of modelers and head and neck oncology researchers. The results demonstrate that our system effectively supports clinical and symptom research.

}


\teaser{
  \centering

  \includegraphics[width=0.99\linewidth]{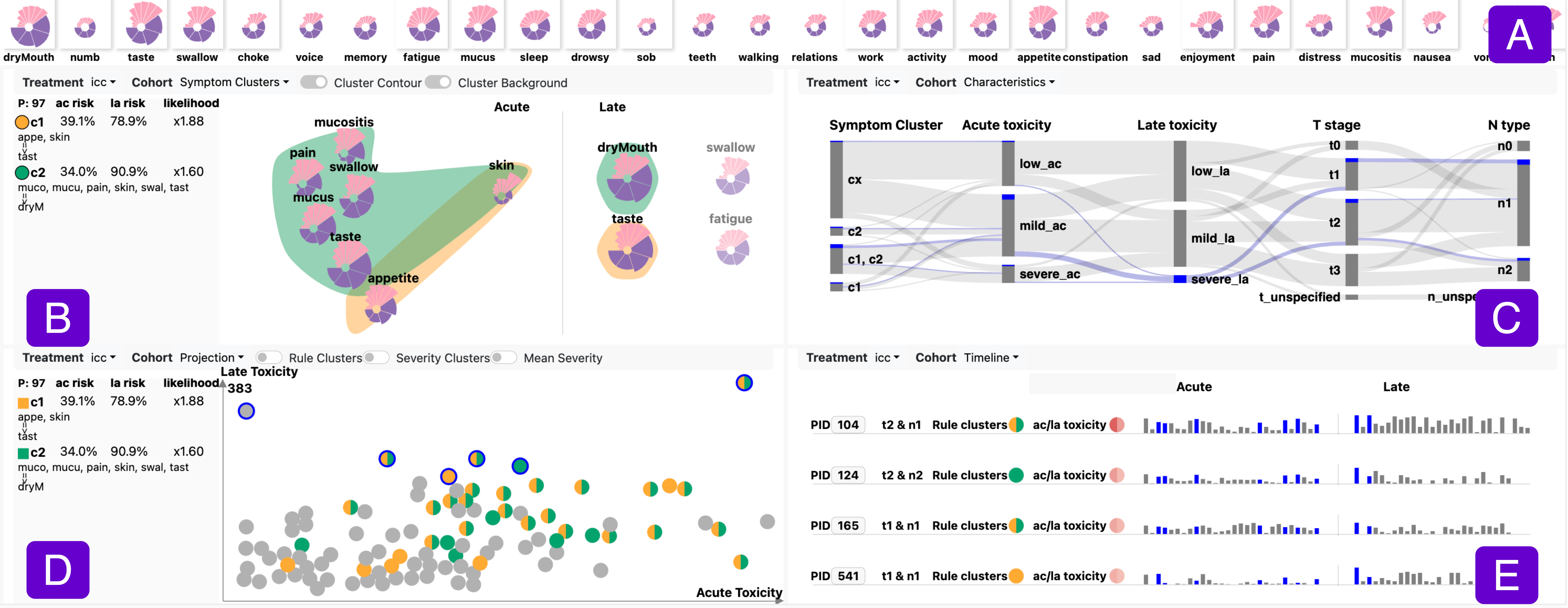}
  \caption{%
  	Longitudinal symptom analysis and prediction for head and neck patients (ICC treatment). A) Overall severity over time for each symptom, across treatments. B) Sequential mining component, showing two clusters that use \textit{acute} symptoms (left) to predict \textit{late} symptoms (right). Lower opacity indicates other \textit{late} prevalent symptoms, not selected by the current model. C) Cohort characteristics, showing symptom cluster results against patient attributes. D) Scatterplot showing patients projected based on the total symptom score for \textit{acute} (X axis) and \textit{late} (Y axis) stages. E) Cohort timeline, displaying cluster labels, clinical details, and mean symptoms burden.
  }
  \label{fig:teaser}
}

\graphicspath{{figs/}{figures/}{pictures/}{images/}{./}} 
\usepackage{tabu}                      
\usepackage{booktabs}                  
\usepackage{lipsum}                    
\usepackage{mwe}                       
\usepackage{mathptmx}                  

\begin{document}


\firstsection{Introduction}
\maketitle
Personalized therapeutics in oncology have resulted in a greater variety of 
 head and neck cancer (HNC) treatment outcomes for patients. Despite the increase in survival outcomes (``roses”), in many patients treatment leads to side effects that can greatly affect quality of life even after the completion of treatment (``thorns”). These symptoms can often be mitigated through preventative therapies, but the preventative treatment can also be an additional burden to patients. Thus, there is a growing  interest in understanding how symptoms develop, stratifying patients into high-risk and low-risk cohorts, and studying the relationship among symptoms and treatment decisions, with a focus on identifying long-term symptoms that affect the patient quality of life.

In HNC, identifying symptom risk is particularly challenging due to the composite effects of specific treatments and clinical factors~\cite{osullivan2016development}. Furthermore, some symptoms are correlated, either due to direct influence, or by shared root causes. These factors make predicting treatment outcomes difficult, and hamper personalized care decision making. Thus, there is a need for alternative human-machine analysis tools that can leverage computational and human effort to help modelers better understand HNC symptoms.

Current computational symptom research is focused on symptom clustering~\cite{gwede2008exploring, kim2013analytical}, however, there is little work~\cite{floricel2021thalis} to explore symptom patterns across time or to compare the outcomes of different treatments. Sequential rule mining (SRM) is a promising unsupervised learning approach for discovering common temporal patterns in the multivariate symptom data, but it can produce many repetitive, or even misleading results for predicting outcomes. Our work uses SRM modeling in combination with other unsupervised machine learning (ML) methods to predict treatment-related toxicities. At the same time, the model results have to also make sense in a clinical setting, and so they need to be interpreted by domain experts. Beyond helping modelers, visual analysis can further help with model interpretation in the context of real clinical patient data.

Visual computing with temporal multivariate symptoms poses several challenges. First, the large size of patient cohort, number of symptoms, and timepoints requires scalable encodings, as well as meaningful aggregation techniques. Second, interpreting symptom trajectories in a clinical setting requires access to clinical features for the cohort. Third, because domain experts are interested in identifying which symptoms are caused by treatments or other symptoms, a visual system needs to allow \textcolor{black}{comparison between symptom groups and between treatments.} Fourth, since the interpretation of \textcolor{black}{association} structures requires both data mining and clinical expertise, the systems need to allow for multiple workflows and levels of details to analyze both symptom and patient sub-cohorts. Last, drawing conclusions from high-dimensional cohort data requires the use of interpretable algorithms, such as rule mining, to help extract patterns that are both useful and simple.

To address these challenges, we introduce a visual computing system to support the analysis and prediction of post treatment symptoms, based on during-treatment symptoms. Our system uses an unsupervised, multivariate method that incorporates sequential rule mining, hierarchical clustering, and factor analysis to assess temporal interrelationships among multiple symptoms in the context of personalized care delivery. Our main contributions are: 1) a description of the modeling problem, data, and tasks; 2) a hybrid human-machine approach for identifying symptom profiles in HNC patients, stratified by treatment methods; 3) the design and implementation of this approach in a system which allows for the exploration of HNC cohort data, with an emphasis on capturing longitudinal patterns in symptom and patient cohorts; 4) a clinically-validated evaluation by domain experts; 5) the lessons learned from this multidisciplinary collaboration.

\section{Related Work}
{\bf Patient Cohort and Clinical Pathway Visualization.} Visual analysis for patient cohorts often relies on finding connections between different patient attributes from medical records. This implies human interpretation of patterns within heterogeneous, and even multidimensional clinical information from patient records. In explainable AI (XAI) medical applications, cohort analysis tackles clinical statistics from patient records~\cite{wongsuphasawat2011outflow, huang2015richly}, cohort history comparison~\cite{zhang2014iterative, chui2011visual, bernard2014visual}, cohort medical image attribute comparison~\cite{maries2013grace, raidou2015visual, klemm2014interactive, wentzel2020explainable,canahuate2023spatially}, or survival risk analysis~\cite{marai2018precision}. The use of visual encodings vary largely among these applications, from custom histograms~\cite{baumgartl2020search}, to time-series plots~\cite{gotz2014decisionflow, jin2020carepre}, matrices~\cite{malik2015cohort, du2016eventaction}, and radial charts~\cite{guo2019visualizing}. When working with large cohorts where the focus is on finding outlier patients and understanding why they are showing unexpected clinical attributes, scatterplot projections are a common way to interpret cohort clusters~\cite{metsalu2015clustvis, meuschke2021gucci, elshehaly2022creative, wentzel2019cohort, floricel2022opening}. Similarly, we use scatterplots for cohort interpretation, however we customize these plots to capture multivariate patient attributes, \textcolor{black}{while also supporting treatment comparison.}

Patient longitudinal medical records data are often visualized using clinical pathway summaries for individual patients~\cite{caballero2017visual}, or cohort temporal summaries for cohorts~\cite{wongsuphasawat2011lifeflow}. Visual abstractions for temporal cohort data have mostly used matrix-based~\cite{du2016eventaction}, flow-based representations~\cite{gotz2014decisionflow, wongsuphasawat2011outflow}, or timelines~\cite{plaisant1996lifelines, baumgartl2020search,guo2017eventthread}. Tree-based representations have been used for event sequence summarization, ordering, and statistics in temporal, clinical data~\cite{malik2015cohort, wang2009temporal, wongsuphasawat2011lifeflow,monroe2013temporal}. Other systems have used PCPs or flow-based representations with line bundling~\cite{meuschke2021gucci, baumgartl2020search}.  \textcolor{black}{While we adapt some of these encodings, we also support cohort summaries of both temporal and categorical attributes.}

A popular method for visual temporal cohort analysis focused on clinical event sequences is sequential pattern mining~\cite{di2020s,cappers2017exploring, wang2021threadstates}. However, sequential pattern mining can be misleading as there is no assessment of the probability that a pattern will be followed. In contrast, our proposed work uses sequential rule mining (SRM), which takes into account the probability that a temporal pattern will be followed. 

\noindent{\bf Rule Visualization.} 
Rule-based modeling is a common approach for creating explainable models~\cite{yang2017scalable, lakkaraju2016interpretable}. In XAI, rule-based explanations are often used to interpret black-box models such as neural networks~\cite{ming2018rulematrix}, support vector machines~\cite{martens2008decompositional}, and latent factor models~\cite{peake2018explanation}. Rules have been adopted in medical data visualization as well, with applications in clinical risk prognosis~\cite{antweiler2022visualizing, benjamin2015interpretable} and disease or treatment toxicity prediction~\cite{ming2018rulematrix,floricel2021thalis,tandan2021discovering,wentzel2023dass}. Surveys and recent visualization systems have shown that rule sets are usually visualized using node-links, tree-based representations, matrices, scatterplots, or PCPs~\cite{bruzzese2008visual,jentner2019visualization, huat2002crystalclear,zhao2018iforest}. \textcolor{black}{In a further departure from previous work, our approach combats scalability issues for large rule sets, and emphasizes the temporal separation between the rule antecedent and consequent.}

Alongside rule sets items, visualization systems have to also integrate relevant rule metrics such as the support and confidence to denote the relevance of the rules. Yuan et al.~\cite{yuan2021exploration} found that feature alignment and predicate encoding are influential visual factors for representing rules, arguing that the interpretability and decision making process are highly influenced by the different rule structures. Applications that support rule itemsets and rule metrics explanation in disease progression have used matrix-based representations accompanied by barcharts and tree-based circular glyphs ~\cite{ming2018rulematrix, antweiler2022visualizing}, while others employed node-links to represent temporal rules from diagnosis codes~\cite{nguyen2018ltarm}. \textcolor{black}{Our previous rule mining work~\cite{floricel2021thalis} explored symptom associations in a given treatment stage independently (\textit{acute} or \textit{late}), and could not capture dependency between stages. In this work we tackle a different modeling problem and XAI challenges, where we focus on late symptoms, we model the symptom burden evolution sequentially, we apply rule clustering to reduce complexity and tackle scalability, and we support per-treatment sub-cohort analysis.}

\section{Background} \label{sec:background}
HNC treatment can involve surgery, radiation, induction chemotherapy, or a combination of these treatments~\cite{tardini2022optimal}. These treatment modalities often result in symptom burden both during the treatment period, and even after the completion of treatment~\cite{eraj2017long}, i.e., “roses have thorns”. The M.D. Anderson Cancer Center documents and quantifies these symptoms through a standardized monitoring program based on MDASI (M.D. Anderson Symptom Inventory)~\cite{Cleeland2000MDASI}, a patient-reported outcome measure for clinical and research use. The program uses questionnaires that are collected weekly at the time of the treatment appointment \textcolor{black}{(acute stage)}, and at longer intervals post treatment \textcolor{black}{(\textit{late} stage), during cancer recurrence monitoring.} 

\textcolor{black}{Counter-intuitively, the two-stage monitoring frequency~\cite{van2021impact} is not driven by current modeling interests: clinicians have a good understanding of symptom values and trends during treatment (\textit{acute} stage), which are sampled with higher-resolution, but not of the after-treatment symptoms (\textit{late} stage), collected at lower-resolution~\cite{wentzel2020precision, christopherson2019chronic}. Data sampling is in fact constrained by the standard of care practice, which is targeted at detecting cancer recurrence~\cite{tosado2020clustering, zdilar2018evaluating} based on recommended guidelines while reducing the patient burden of required clinic visits, and the fact that in-clinic questionnaires yield higher reliability and patient compliance than at home self-reports.} 

Since cancer patients can experience a multitude of symptoms that can co-occur or can cause other symptoms, oncologists are interested in modeling clusters of frequently co-occurring symptoms and in how symptoms are correlated with the diagnosis~\cite{wentzel2021precision,wang2021predicting} and prescribed treatment~\cite{aktas2010symptom, dong2016symptom, fan2007symptom,wentzel2023dass,vandijk2023head}. However, existing research does not focus on the temporal \textcolor{black}{association} between symptoms, changes in symptom severity over time, or the prediction of post-treatment symptoms. Current approaches that analyze the \textcolor{black}{association} between symptoms include methods such as factor analysis (FA)~\cite{rosenthal2007measuring, skerman2009multivariate}, \textcolor{black}{hierarchical cluster analysis} (HCA)~\cite{gwede2008exploring}, latent class profile analysis~\cite{illi2012association,kim2013analytical}, and rule mining~\cite{biggs2021association, floricel2021thalis}. In this work, we study how \textit{acute} symptoms predict \textit{late} symptoms, using a combination of SRM and HCA.

\section{Design}
This project is part of a multiyear, interdisciplinary collaboration between research groups  with cancer symptom modeling experience from three research sites, composed of: three radiation oncology experts with clinical and research experience, one senior data mining expert, one senior visual computing expert, and several junior researchers in visual computing. The team held weekly remote meetings to discuss various clinical data analyses, during which our visual computing research group collected feedback.

Our design process followed an Activity-Centered-Design (ACD) approach~\cite{marai2017activity}, focusing on user activities and workflows. This paradigm has shown higher success than traditional human centered design for scientific, interdisciplinary collaborations. In this work, we used ACD to build workflows around the evaluation of clinically applicable models and complementary clinical data analysis.

The visual computing and data mining research groups met weekly to define functional specifications, prototype the interface for the clinically applied models, and evaluate the interface. This was an interactive process that, following the ACD approach, proved to be effective in the context of this remote collaboration~\cite{marai2017activity}. Alternative designs for visual prototypes are available in the supplemental materials.

\subsection{Activity and Task Analysis}
Our system serves model builders in cancer symptom research. Our collaborators have experience in ML approaches for symptom analysis, but were interested in alternative approaches for temporal analysis that are centered on exploring the differences between patients receiving different treatment modalities, \textcolor{black}{with particular emphasis on \textit{late} symptoms}. There was also a need to efficiently present and interpret the results from the proposed model to our clinician collaborators. Additionally, it was imperative to compare the toxicities found by the model across different treatment groups. Based on these considerations, and following the ACD paradigm, we split the requirements for this project into two main activities and we list their corresponding tasks:

\noindent\textbf{A1} Support temporal symptom analysis for a given treatment
    \begin{itemize}
        \item \vspace{-0.3em} \textbf{T1.1} Predict \textit{late} symptoms based on \textit{acute} symptoms
        \item \vspace{-0.3em} \textbf{T1.2} Identify temporal patterns in the overall symptom severity
        \item \vspace{-0.3em} \textbf{T1.3} Correlate clinical cohort details and symptom patterns
        \item \vspace{-0.3em} \textbf{T1.4} Facilitate the analysis of a subset of patients within a cohort
    \end{itemize}
\textbf{A2} Support temporal symptom analysis across multiple treatments
    \begin{itemize} 
        \item \vspace{-0.3em} \textbf{T2.1} Compare temporal symptom profiles across treatments
        \item \vspace{-0.3em} \textbf{T2.2} Evaluate the likelihood of experiencing a symptom profile compared to alternative treatments
        \item \vspace{-0.3em} \textbf{T2.3} Identify temporal patterns in severity across treatments
        \item \vspace{-0.3em} \textbf{T2.4} Facilitate the comparison of clinical patient data for multiple treatments
    \end{itemize}

Our evaluation describes examples of preferred workflows concentrated on these activities, while the results are clinically validated by oncology domain experts. Non-functional requirements included clarity in the model results, scalable visualizations that can display symptom and patient statistics, and intuitive visual abstractions.

\begin{figure*} 
\centering
\includegraphics[width=0.9\linewidth]{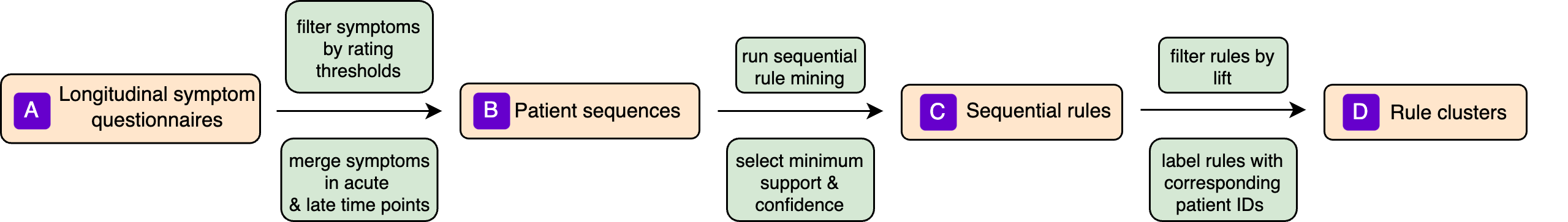}
\caption{SRM Modeling. A) Patient-reported symptom ratings are recorded as longitudinal records. B) Records are processed into patient symptom sequences. C) Patient sequences are provided as input to the SRM algorithm. D) The sequential rules are filtered and clustered into rule clusters based on their corresponding patient IDs.
}
\label{fig:fig2}
\end{figure*}

\subsection{Data}
Our data is from a cohort of 823 HNC patients who underwent treatment at the M.D. Anderson Cancer Center in Houston, TX. Demographic and diagnostic information was recorded for this cohort, spanning ordinal attributes (tumor stage, lymph node stage), quantitative attributes (age, radiation dose), nominal attributes (treatment modality), and time-series attributes with quantitative values (symptom ratings) collected in two stages, \textit{acute} (baseline and during treatment, higher frequency) and \textit{late} (after treatment, lower frequency). 

Self-reported longitudinal (temporal) multivariate  symptom data was extracted from patient questionnaires~\cite{Cleeland2000MDASI} over the span of 12 time points: before starting the treatment, weekly for 7 weeks during treatment, 6 weeks after treatment, and 6, 12, and 18 months post treatment. Symptoms were rated on a 0-to-10 scale, from "not present” (0), to "as bad as you can imagine'' (10). A total of 28 symptom variables were considered in this longitudinal assessment, split into HNC specific symptoms (swallow, speech, mucus, taste, constipation, teeth, mouth sores, choking, and skin problems), general cancer symptoms (fatigue, sleep, distress, pain, drowsiness, sadness, memory, numbness, dry mouth, appetite, breath, nausea, and vomiting problems), and daily life interference symptoms (work, enjoyment, general activity, mood, walking, relationships problems). The 12 timepoints belong to one of two categories: the \textit{acute} stage (once before the treatment’s start date, or week 0, and all 7 weeks throughout the treatment), and the \textit{late} stage (the remaining 4 post treatment assessment dates). Not all features were available for every patient. Missing clinical variables were marked as “unspecified”, and missing symptom ratings were considered a rating of 0, which were not considered when building the models.

This cohort presents six possible treatment combinations: induction with concurrent chemotherapy and radiation therapy (ICC) (n = 97), concurrent chemotherapy and radiation therapy(CC) (n = 329), induction and radiation therapy (IRT) (n = 66), radiation therapy alone (RT) (n = 199), surgery and other treatments (S\_and\_others) (n = 75), and surgery alone (S) (n=57). Patients were stratified by treatment during the sequential rule mining analyses. Patients receiving surgery alone were removed from the model building because this sub-cohort did not report weekly symptom scores during treatment.

\subsection{SRM Modeling for Medical Data}
Association Rule Mining (ARM)~\cite{agrawal1994fast} is an unsupervised method that identifies frequent patterns, correlations, or \textcolor{black}{association} structures in transactional data sets. Association rules are most commonly found in the form $X \rightarrow Y$ (the appearance of X implies the appearance of Y), with X called the antecedent and Y the consequent of the rule. Because rule mining is more transparent than the black box models used in diverse applications, it has caught attention in medical research as well~\cite{antweiler2022visualizing,tandan2021discovering,nguyen2018ltarm}. We applied ARM in our previous work~\cite{biggs2021association,floricel2021thalis} in the context of cancer symptoms $\{taste\}\rightarrow \{dryMouth\}$ (if the patient suffers from taste, then they will \textcolor{black}{more} likely suffer from dryMouth as well) by transforming our longitudinal symptom records into a transactional data set. This helped to find common symptom combinations at different stages in the patient observation period, but it did not help us predict \textit{late} symptoms based on symptoms during treatment.

One interesting extension of association mining for temporal data is sequential rule mining (SRM)~\cite{deogun2005prediction}. SRM uses the antecedent of a rule to predict the consequent of the rule with the condition that the antecedent precedes the consequent. We applied SRM to our longitudinal symptom data considering the during- and post-treatment time frames as temporal sequences of symptom toxicity as follows:

 \begin{equation}
 \centering
      R1: \{taste,nausea\}\rightarrow \{dryMouth\}
 \end{equation}

\noindent meaning that if a patient suffers from taste and nausea problems during treatment, they will \textcolor{black}{more} likely suffer from dryMouth problems after the completion of the treatment. However, the \textcolor{black}{ disadvantage of rule mining in clinical applications is that typically a large number of rules may be required to make knowledge actionable.
} Moreover, prediction should reflect a strong \textcolor{black}{association} relationship between the antecedent and the consequent of a rule. Fortunately, useful knowledge can be quickly identified using rule metrics such as support, confidence, and lift. In the case of the previous rule $R1$, the support of the rule is the ratio of patients that have taste and nausea problems during treatment followed by dryMouth problems after treatment:

\begin{equation}
 \centering
 \resizebox{0.6\linewidth}{!}{
${sup(R1) = \frac{|\{(taste,nausea) \cup (dryMouth)\}|}{|S|}}$
}
\end{equation}

\noindent
where $|S|$ is the total number of patient symptom sequences.

The confidence of the rule predicts the risk of a patient to develop \textit{late} symptoms (dryMouth in our example), given a certain symptomatology during treatment (taste and nausea in our example) and is reported as: 

\begin{equation}
 \centering
 \resizebox{0.5\linewidth}{!}{
${conf(R1) = \frac{sup(R1)}{sup(\{taste, nausea\})}}$
}
\end{equation}

The lift of a sequential rule denotes the strength of the rule, or in other words, denotes whether the antecedent and the consequent are dependent on each other or not, and is computed as follows:

\begin{equation}
 \centering
 \resizebox{0.7\linewidth}{!}{
${lift(R1) = \frac{sup(R1)}{sup(\{taste, nausea\}) \times sup(\{dryMouth\})}}$
}
\end{equation}

A lift value $\le$1 indicates that the rule is not able to predict the consequent more accurately than what could be predicted by chance. 

As already noted, rule mining can result in a multitude of rules that can show overlapping patterns. It is important to filter these results based on the previous metrics in order to get useful, easy to interpret, and meaningful information regarding the patterns within the data.

\subsubsection{Back-end Design} \label{sec:srm}
We use Sequential Rule Mining (SRM) to identify temporal patterns in symptoms and to predict \textit{late} symptoms. We discretize treatment ratings into two bins: before treatment and weekly ratings taken during treatment for up to 7 weeks (the \textit{acute} stage), and ratings 6-18 months after treatment (the \textit{late} stage) (\Cref{fig:fig2}.A). Patients are stratified based on treatment modality, and the rule mining algorithm is run separately for each sub-cohort, as we are interested in identifying treatment-related symptoms.

We used the CMDeo algorithm~\cite{fournier2012cmrules} to compute the sequential rules, which is an adaptation from Deogun et al.’s algorithm~\cite{deogun2005prediction} for multiple sequences of events. We followed the documentation from the open source data mining library called SPMF~\cite{fournier2014spmf} that supports the CMDeo algorithm. The Python wrapper from this library was used for the model, which required us to pre-process our data to correspond to the input structure from the documentation. 

In the first step of the data pre-processing, we computed sequences from the patient timelines (\Cref{fig:fig2}.B). 
 \textcolor{black}{Each sequence corresponds to the temporal ratings of one patient across both the \textit{acute} stage (baseline and during treatment) and the \textit{late} stage (after treatment). Accordingly, we abstract the sequences into two-stage patterns, \textit{acute} and \textit{late} (Sec.~\ref{sec:background}). In the \textit{acute} pattern, we include a symptom only if the patient provided a rating above a given severity threshold (e.g. $\ge$5) during any of the \textit{acute} time points. Similarly, in the \textit{late} pattern, we include a symptom only if the patient provided a rating above a given threshold (e.g. $\ge$3).} Clinically, a rating $\ge$5 is considered a moderate-to-high severity, while 3 is considered mild severity. The same threshold is not enforce for the two stages because in general, ratings are lower in the \textit{late} stage than in the \textit{acute} stage. The use of a severity threshold helps to minimize patient variability and individual symptom severity ratings.

Next, the SRM algorithm was applied on these sequences to identify sequential rules (\Cref{fig:fig2}.C). Similarly to traditional association rule mining, two input parameters, namely support and confidence, need to be specified by the user to generate the rules. In our experiments, we used minimum support (i.e. percent of patients that show the resulting patterns) of 30\% or 40\% depending on the number of sequences, as we consider patterns experienced by a third of the patients to be significant. The minimum confidence (i.e. risk of \textit{late} symptoms) was set to 50\%. From the initial set of rules, only rules with a lift threshold higher than 1 were selected to ensure the rules can be used for the prediction of \textit{late} symptoms.  
The lift of a rule indicates the degree of dependency between the antecedent and consequent of the rule. The resulting rule sets varied from 9 to 46 rules, depending on the number of sequences for each treatment and the variety of occurring symptoms per sub-cohort.

As could be expected, the extracted rules within each treatment cohort showed a lot of similarities in terms of the symptom patterns (often differing in only one or two symptoms) and in the set of patients supporting the rules (over 90\% of the same patients appearing in two or more rules). To minimize redundancy among the rules, we decided to cluster the rules into rule clusters that would then be used for visualization. We labeled each rule with the corresponding patient IDs supporting the rule. Next, we computed the similarity between rules based on their common patient IDs using Jaccard’s index ~\cite{jaccard1912distribution}. We used this method because we work with sets (i.e. patient ID sets) for which we wish to compute rule similarity based on the patients the rules affect. We then applied hierarchical clustering using the complete linkage~\cite{defays1977efficient} on the resulting similarity matrices. We used the complete linkage since \textcolor{black}{the point of reducing a group of rules to a single rule was to yield cohesive rule clusters while avoiding in-cluster outliers}. We used hierarchical clustering since we have found it yields highly interpretable results through the use of dendrograms~\cite{luciani2020spatial} which allows us to manually adjust the clusters and identify outliers. We decided upon the number of clusters after inspecting all treatment results. We created rule clusters (\cref{fig:fig2}.D) by merging the antecedent symptoms and consequent symptoms from all rules within a cluster. Thus, each cluster is formed by a set of \textit{acute} symptoms and a set of \textit{late} symptoms.

We attached to each cluster all the patient IDs from that cluster’s corresponding rules. This is helpful for visually connecting the cluster information with the patient cohort. We report the following measurements per each cluster: 1) the probability (support) of developing the \textit{acute} symptoms given a treatment method; 2) the probability of the \textit{acute} symptoms to develop the cluster’s corresponding \textit{late} symptoms, given by the confidence of the rule cluster; 3) the likelihood that the temporal pattern shown by the rule cluster will appear more frequently as compared to the rest of the treatment modalities, given by the support of the cluster within the treatment over the support of the cluster outside the treatment (i.e. for all the alternative treatment modalities).

\subsection{Front-end Design}
Our system uses Python for the back-end and React with D3.js for the front-end. The top design is based on coordinated multiple views, which support diverse analysis workflows. The interface consists of 5 panels that support 6 types of visual components. The top panel (\Cref{fig:teaser}.A) serves deliberately as an anchor which cannot be configured by the modeler, and displays the stratified overall symptom severity for the entire patient cohort. The remaining quadrants can be configured with any of the following five visual components: symptom clustering  (\Cref{fig:teaser}.B) - which denotes temporal symptom clusters for one treatment; patient clustering  (\Cref{fig:teaser}.D) - which shows patient cohort symptomatology attributes for one treatment; cohort characteristics (\Cref{fig:teaser}.C) - which correlates diagnostic data to symptom clusters and symptom overall severity over time; cohort timeline (\Cref{fig:teaser}.E) - which displays an in-depth view of each patient's longitudinal and diagnostic features; and the symptom query component (\Cref{fig:fig5}.C) - which provides overall statistics regarding the appearance of symptoms during (\textit{acute}) and after treatment (\textit{late}). Excluding the top view, which uses the entire cohort, each component displays the data for one treatment modality. The quadrants have treatment and visual component queries attached to their top-left to facilitate workflow configurations. \textcolor{black}{This front-end design supports our modeling goals, which are centered on treatment comparison and cluster outlier analysis, but not on patient comparison, nor on alternative symptom rankings.}

\begin{figure}[t!]
\centering
\includegraphics[width=0.5\linewidth]{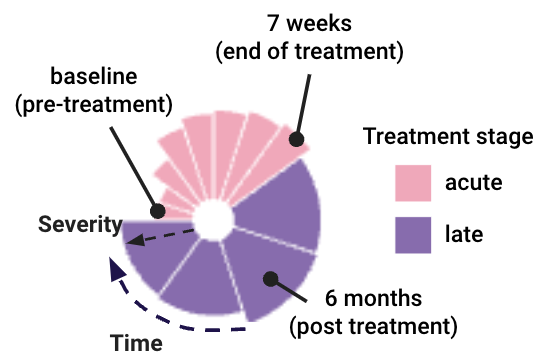}
\caption{Rose glyph. Color-coded petals aggregate the mean severity for patients for symptom dryMouth. Petals in the radial layout start at 9 o'clock and proceed clockwise. Pink “petals” encode \textit{acute} time \textcolor{black}{intervals} while purple encodes \textit{late} time \textcolor{black}{intervals}. \textit{Late} \textcolor{black}{ petals are wider to depict longer time intervals, while \textit{acute} petals depict shorter intervals.}
}
 \label{fig:fig3}
\end{figure}

Our design uses custom Rose Glyphs (\Cref{fig:fig3}) to encode the trajectory of a single symptom severity within either the entire cohort (\Cref{fig:teaser}.A), or subgroups in the data (\Cref{fig:teaser}.B). For the selected subgroup, the mean symptom rating at each time interval is encoded using variable-radius slices (petals). The symptom trajectory starts with the baseline ratings at 9 o’clock, and progresses clockwise in order of increasing time intervals, showing rating details for each of the twelve time intervals of the patient observation period. Pink petals encode \textit{acute} treatment time intervals, while purple petals encode \textit{late} intervals. \textcolor{black}{The width of the petals is driven by the two-stage data sampling, and by the modelers’ interests in late symptoms (Sec.~\ref{sec:background}). The color scheme was chosen based on perceptual criteria~\cite{marai2019ten}, and to emphasize late intervals. The flat interval mapping was in alignment with the clinical assumption that symptom variation within an observation interval is not significant. } We took inspiration from Florence Nightingale's Rose Diagram~\cite{brasseur2005florence} for this glyph, but instead of focusing on comparing events within a time frame, we concentrate on temporal trajectory and comparing trajectories across symptoms. We employed the radial glyph design because it provided a compact way to display symptom \textcolor{black}{burden across cohorts and treatments, while supporting rapid similarity detection~\cite{marai2018precision}. Our timeline histogram supports symptom value comparison better than the glyph layout; however, the glyph’s purpose is not accurate symptom value comparison, but compactly encoding each symptom’s trajectory and comparing symptom trajectories.}

\subsubsection{Overall Symptom Severity}
This component (\Cref{fig:teaser}.A) displays the mean severity (i.e. rating) distribution for each of the 28 symptoms for the entire patient cohort (i.e. all treatment modalities) (T2.3) using rose glyphs. The symptom list starts with dryMouth, which is the one of the most severe symptoms throughout the observation period, and it is the most persistent symptom after treatment across patient sub-cohorts. The rest of the symptoms are ordered based on the cosine similarity to dryMouth, computed using the mean temporal ratings per each symptom. We used cosine similarity because we are more interested in relative frequency of symptom occurrence, as there can be a large variation in self-reported symptoms among items that may not correlate to their impact on quality-of-life. \textcolor{black}{The symptoms are grouped based on temporal similarity, in support of our modeling goals. While other grouping options are possible, they were g priority in this project.} Symptoms predicted by SRM in at least one of the treatments are highlighted with a shadowed border. This encoding provides a compact way of showing overall symptom burden for the entire cohort and it serves as a reference point for evaluating treatment-specific symptom patterns.

\subsubsection{Symptom Clustering}
\begin{figure}[t!]
\centering
\includegraphics[width=0.95\linewidth]{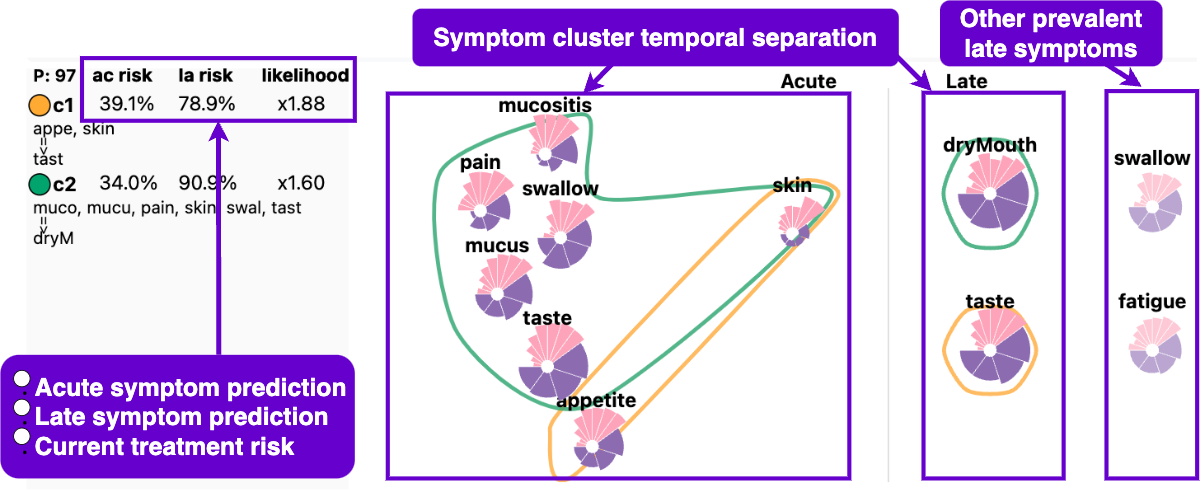}
\caption{ Symptom clustering for treatment ICC. The clusters in orange and green predict in the \textit{late} stage dryMouth and taste problems. Cluster 1 (orange) shows a higher risk to develop these toxicities for ICC rather than the other existing treatment modalities (i.e. 1.88 times more likely).
}
 \label{fig:fig8}
\end{figure}

This component (\Cref{fig:fig8}) provides a visual abstraction for the symptom clusters found in \Cref{sec:srm} through a 2D projection of the corresponding symptoms using rose glyphs (T1.1, T2.1). The view is split into two halves to facilitate the temporal separation between \textit{acute} and \textit{late} stages. The X and Y axis in the \textit{acute} half correspond to the first two principal components after applying PCA on the Jaccard’s similarity between symptoms, based on the common patient IDs they share. Because many of these symptoms have underlying association, we used PCA, as opposed to other projections, as it works better for correlated attributes. We use a force-directed layout to ensure the symptom glyphs are not overlapping in the projection. Symptoms are represented using rose glyphs to show the mean severity distribution over time across the patients that correspond the clusters. This also enhances temporal symptom severity comparisons between a selected treatment and the overall cohort or another treatment.

In the \textit{late} half (\Cref{fig:fig8}), the clustering results are not part of the PCA projections because these clusters usually resulted in one or two different symptoms in this stage for a given treatment. Additionally, we listed on the right edge of the view the \textit{late} symptoms that appeared in our rule mining results, but were not part of the rules filtered for the prediction or the clustering of the symptoms due to low metrics results (i.e. lift < 1). We chose to visualize these additional \textit{late} symptoms to highlight the fact that, although the data shows many common treatment-related toxicities, these cannot be accurately predicted using \textit{acute} symptoms with the data at hand. We mark these symptoms using a low opacity for the rose glyphs as opposed to the predicted symptoms.

The left legend of the component shows the details for each symptom cluster (\Cref{fig:fig8}): the cluster ID, the corresponding antecedent (\textit{acute} symptoms) and consequent (\textit{late} symptoms), he support of the \textit{acute} stage (i.e. how many patients display the symptom patterns from the \textit{acute} stage), the confidence of the cluster (i.e. the risk of developing \textit{late} symptoms given the \textit{acute} symptoms), and the support of the cluster within the treatment cohort over the support of the cluster for the other treatment cohorts (i.e. the likeliness that this cluster might appear more frequent for the given treatment as opposed to all the other treatments) (T2.2). Each cluster is highlighted using an envelope (\Cref{fig:fig5}.A,B) categorically color-coded~\cite{Wenskovitch2014}. The envelopes’ background can be turned on (\Cref{fig:teaser}.B), which can better emphasize the symptom correspondence to clusters, using a Venn diagram-like illustration. From the legend panel, the clusters can be unselected, which will result in the removal of the highlight for those cluster envelopes. Selecting a symptom glyph from the projection or a cluster label from the legend will result in highlighting the complementing information in the other interface components (e.g. cohort attributes that correspond to the selected item).

In our previous work with rule mining for symptom analysis, we used node-link diagrams to represent the symptoms’ inter-relationships [21]. Domain experts preferred this 2D visual abstraction due to the small number of rules that we displayed. However, in this project we work with a larger number of temporal rules. Early prototypes relied on a combination of network-based encodings and barplots. However, this resulted in clutter due to the large number of edges between nodes that did not capture well the temporal nature of the rulesets. As a result, we detached from displaying actual rules and opted for a cleaner pro- jection that uses rule clusters, using envelopes to show relationships between symptoms and horizontal separation to denote temporal direc- tion. We opted for the rose glyph, as opposed to circles, for symptom interpretation, to enhance trajectory comparison between symptoms.

\begin{figure}[t!]
\centering
\includegraphics[width=\linewidth]{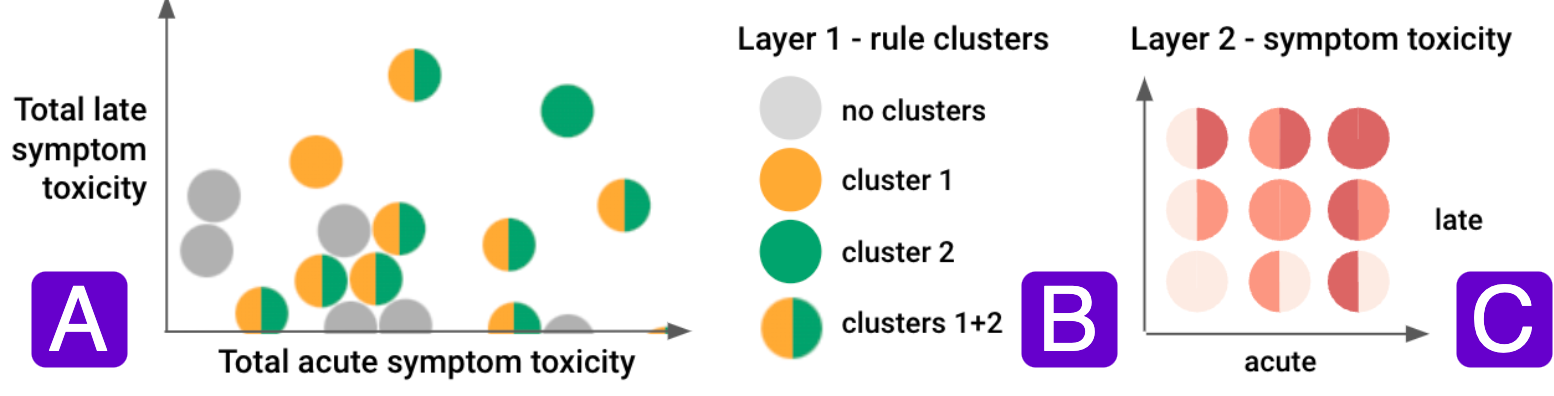}
\caption{Patient Clustering View. A) Scatterplot showing patient glyphs. Two options for patient encodings in the scatterplot: B) encodes cluster memberships and C) encodes temporal symptom burdens.}
 \label{fig:fig4}
\end{figure}

\subsubsection{Cohort Symptom Query}
This component (\Cref{fig:fig5}.C) provides an overview of all the 28 symptoms from the cohort for the \textit{acute} and \textit{late} stages, and guides the analysis of symptom clusters, using a vertical barchart (T2.1). For a selected treatment, tumor and lymph node stage, \textit{acute} and \textit{late} symptom rating thresholds; this view returns the percentage of the patients that have reported symptoms above the given thresholds at least once during and after treatment for each symptom. Symptoms are ordered from top to bottom by the highest cumulative percentages for \textit{acute} and \textit{late} occurrences, which highlights symptoms of high prevalence among patients. Symptoms from SRM clusters are colored in blue.

We proposed this encoding in early prototyping iterations, to show statistics for rule symptom occurrences. Our collaborators quickly adopted it into their analysis due to its low complexity, so we chose to follow this design for explaining symptom prevalence.

\subsubsection{Patient Clustering}
This component (\Cref{fig:teaser}.D) provides a custom 2D scatterplot projection of the patient cohort, with axes corresponding to the total symptom severity scores for the \textit{acute} time points (X axis) and the \textit{late} time points (Y axis) (\Cref{fig:fig4}.A). We chose this orientation to better highlight patient outliers for the \textit{acute} and \textit{late} stages (T1.2). We use a force-directed layout to remove overlap and ensure that each individual patient can be selected from this projection for further analysis. \textcolor{black}{Patient similarity comparison is supported by the scatterplot projections.}

This component has two interchangeable layers: the first layer (\Cref{fig:fig4}.B) colors the points based on the patients’ rule cluster labels. If a patient is not included in any of the rule clusters, their corresponding point is gray. Otherwise the point is split into as many sections as the number of clusters it belongs to, where each section is colored to match the clusters colors from the symptom clustering component. The second layer (\Cref{fig:fig4}.C) splits the points into two sections, representing, from left to right, the \textit{acute} and \textit{late} treatment periods, respectively. The color of each section is mapped to the overall symptom severity for its corresponding period, with lighter red encoding low severity and dark red encoding high severity. This layer can be applied upon selecting a subset of symptoms from the top rose glyph row (\Cref{fig:teaser}.A), and it will be updated to show the \textit{acute}/\textit{late} severities of the selected symptoms. Brushing operations will highlight or filter information in the rest of the views based on the patient selection (T1.4). 

Alternative designs experimented with other projection methods and glyph encodings. However, we found most projection methods like PCA~\cite{floricel2021thalis} and T-SNE did not capture the rule clusters and associations. In contrast, we found moderate-to-high \textit{acute} and \textit{late} symptom ratings were consistently correlated with more cluster membership, which made the glyph encoding more intuitive to collaborators. Using symptom severity made it easier for collaborators to identify patients with increases or decreases in treatment severity between the \textit{acute} and \textit{late} stage. For the scatterplot glyph design, we considered alternative shapes instead of circles for different clusters, but we found that it was difficult to capture an arbitrary number of cluster memberships across treatment modalities using shape. For the symptom toxicity layer, we considered splitting circles into more than two time periods (i.e. \textit{baseline}, \textit{acute}, \textit{late}) or using rose glyphs, but that cluttered the view and made it difficult to find patterns. This component ensures a better understanding of the model results and clinical statistics as it connects that cohort information to actual patients for the given treatment.

\begin{figure}
\centering
\includegraphics[width=0.95\linewidth]{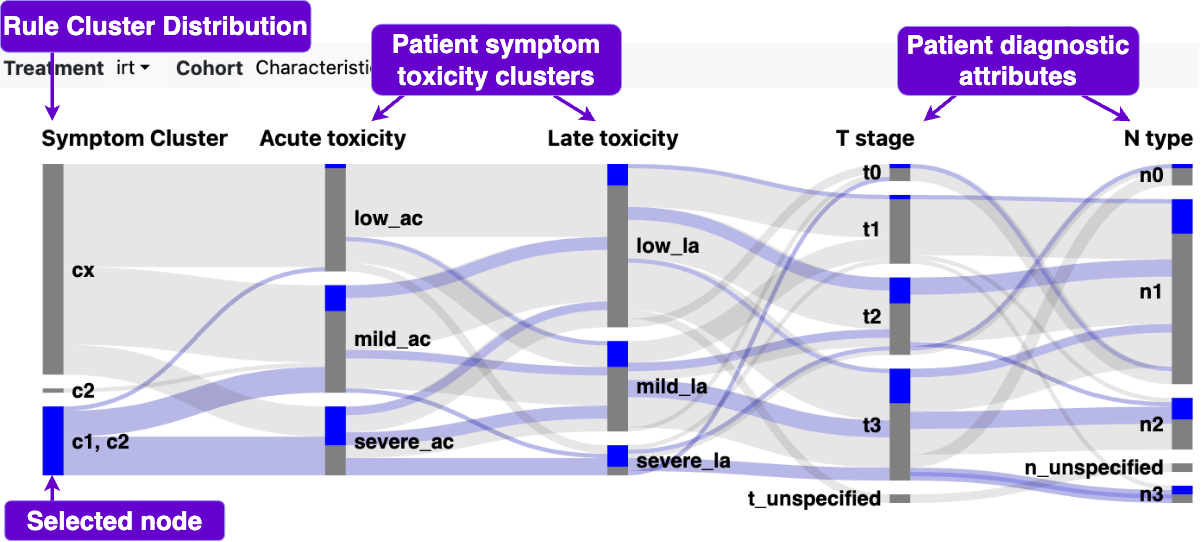}
\caption{Sankey Diagram for IRT treatment. Node $c1,c2$ is selected, showing that a very small part of the patient cohort from with this cluster combination is linked to low symptom severity in the \textit{acute} stage. }
 \label{fig:fig7}
\end{figure}

\begin{figure*} 
\centering
\includegraphics[width=\linewidth]{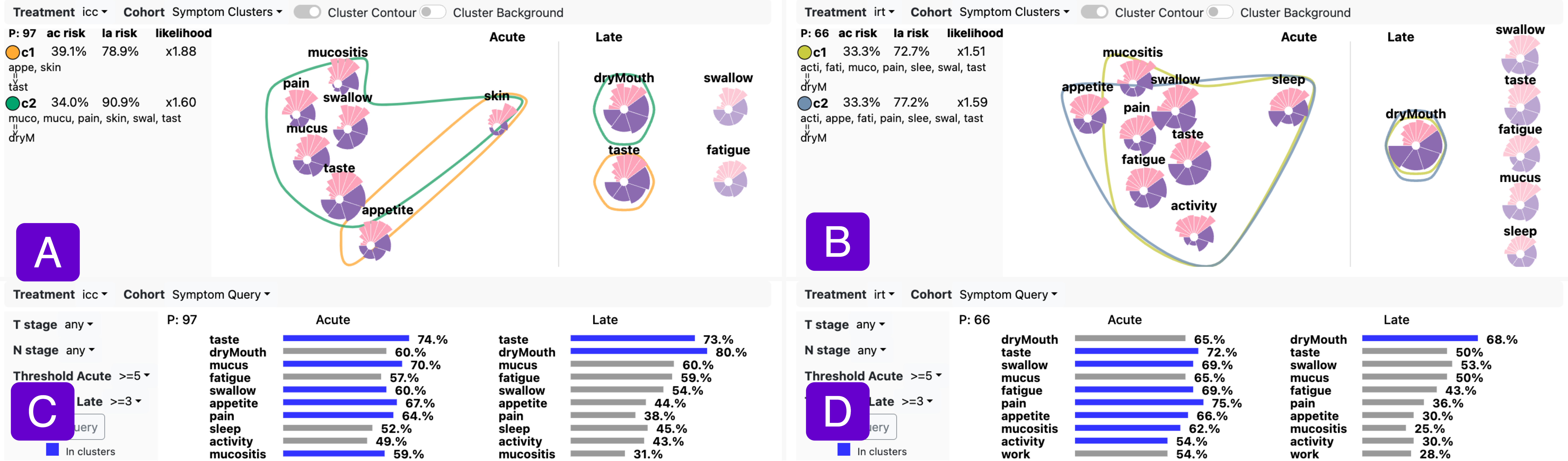}
\caption{Treatment comparison. A) Overall cohort toxicities for all timepoints. B) and C) Symptom clusters for treatments ICC and IRT. Both treatments show two clusters, with similar \textit{acute} symptoms, but ICC presents taste as a \textit{late} symptom (B), as opposed to IRT(C). Although the rose glyphs are projected based on similar patients in the \textit{acute} half, both treatments have outliers (i.e. skin and sleep in \textit{acute}). D) and E) Symptom queries showing the prevalence of all symptoms for the two selected treatments. These bottom views show that, although there are many \textit{late} common toxicities, not all can be predicted by the \textit{acute} symptoms in B) and C) (i.e. mucus in \textit{late} ICC is prevalent but not predicted in the symptom clusters).
}
\label{fig:fig5}
\end{figure*}

\subsubsection{Cohort Timeline}
This component (Figure 1.E) functions as a detailed view of the at- tributes of each patient (T1.4), using timelines and small multiples to show mean symptom ratings over time, patient cluster labels, and diag- nostic information (T1.3). The left half of the view shows the patient’s ID, tumor (T) stage, lymph node (N) stage, symptom clusters labels, and temporal symptom severity using their corresponding points from the scatterplot (Figure 8.C). The right half uses a barchart timeline, split by the \textit{acute} and \textit{late} stages, showcasing mean ratings for each of the 28 symptoms (Figure 8.C). The symptom bars are ordered \textcolor{black}{by} the top interface row symptom order. They are colored in blue when they represent symptoms that are present in at least one of the rule clusters for the selected treatment, or in gray vice versa. \textcolor{black}{The patient timelines are listed in descending order based on the cumulative \textit{acute} and \textit{late} symptom severity, and based on symptom cluster membership in case of equality for the former metric.} Brushing from the scatterplots will filter this view by the selection. Clicking on the patient IDs will highlight the corresponding patients in the scatterplot and in flows in the cohort characteristics component.

Oncology experts are often interested in analyzing a single patient~\cite{sheu2017conditional, luciani2018details} and comparing them to the rest of the cohort. As a result, we designed this component to make individual-patient analysis possible. Previous prototyping iterations explored matrix-based encodings which included all timepoints from the symptom data. This resulted in cluttered components which took the majority of screen space due to the large number of timepoints, making the inclusion of diagnostic patient data difficult. Thus, we adopted this custom, simplified view of the temporal symptom data, deciding to aggregate the \textit{acute} and \textit{late} time points while also integrating the diagnostic and symptom cluster/severity labels. The timeline component can also be used to observe how a patient’s symptomatology trajectory compares to other patients, or to observe the overall burden of symptoms for a given set of patients (T1.2, T2.3).

\subsubsection{Cohort Characteristics}
This component (\Cref{fig:fig7}) connects symptom cluster memberships, overall symptom burden for the \textit{acute} and \textit{late} stages, and diagnostic patient data (T stage, N stage) using a Sankey Diagram (T1.3, T2.4). Apart from showcasing the possible symptom cluster combinations, we stratify the patients into low, medium, or high symptom burden for the \textit{acute} and \textit{late} stages using K Means clustering on the total symptom toxicity scores for both stages. This further emphasizes how \textit{acute} symptom burden transposes into \textit{late} symptom burden. The nodes from the diagram can be selected and the corresponding nodes and flows are highlighted in blue (\Cref{fig:fig7}), while filtering options in the other views highlight with blue the selection in this component as well.

When we prototyped this component, we kept in mind that we needed to showcase the distribution of categorical cohort attributes while also considering time directionality for our temporal attributes (i.e. \textit{acute} and \textit{late} symptom toxicity). We opted for a Sankey design as it has shown adoption in both categorical and temporal attributes in previous work~\cite{wongsuphasawat2011outflow}. This design was easily adopted by our collaborators and became a pivotal component in their analyses. All of the diagram’s ordinal axes are ordered from top to bottom (i.e. T/N stage, \textit{acute}/\textit{late} toxicity), as per the suggestion of our oncology domain experts. Due to the limited number of attributes, this component can clearly show the distribution of a particular attribute’s values for a treatment modality and how it is connected to the distributions of the other cohort attributes.

\subsubsection{Flexible Workflow Support}
Due to the variation of requirements that would support analysis at both the patient cohort level, as well as at the symptom cohort level, we designed these visual components to provide a balance between flexibility and guidance across analysis workflows. Our modelers were interested in understanding and interpreting the SRM model results in the context of treatment decision making and treatment-related symptoms. However, they were also looking for common symptoms across treatments that may develop independent of treatment strategy. Moreover, they were interested in predicting what a new patient should expect given a selected treatment to better assist future treatment decisions. To support these workflows, the afferent components can be flexibly swapped.

\section{Evaluation}
We evaluated the system and the resulting models through multiple demonstrations and case studies involving two senior model builders, two junior model builders, and three senior clinical oncology co-modelers with ML experience. \textcolor{black}{Beyond the six domain expert co-authors of this paper who provided feedback, an additional oncology researcher also evaluated the system.} \textcolor{black}{The model builders were active co-designers, whereas the oncology experts provided occasional input and feedback.} Although our system is dedicated to model builders in cancer symptom research, we also needed to clinically validate the results we had found. \textcolor{black}{The evaluators participated in several hour-long demo sessions over 4 months, followed by the case studies. They also explored the tool on their own while providing feedback.} We illustrate two case studies that were conducted through focus groups via Zoom, using screen sharing and note taking. During these sessions, the first author navigated the interface under the guidance of the model builders and oncology co-modelers, using the think-aloud method. These studies used a cohort of 766 HNC patients that presented five treatment modalities: RT, IRT, CC, ICC, and Surgery\_and\_other. We present below, in abbreviated form, these case studies.

\begin{figure*} 
\centering
\includegraphics[width=\linewidth]{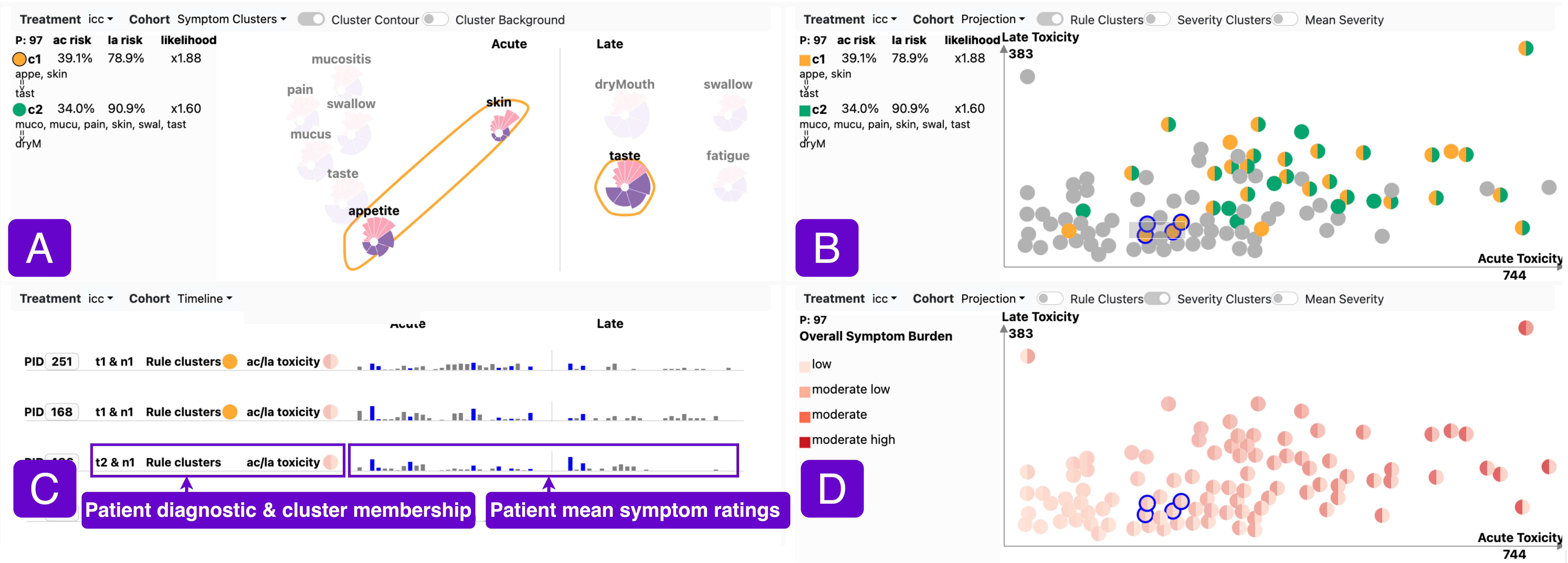}
\caption{Single-treatment analysis. A) ICC treatment symptom clusters with cluster 1 (orange) selected. B) ICC patient projection with the cluster label layer. The cluster 1 outlier patients from the lower-left side are selected and highlighted with blue in the scatterplot filtered in the other views. C) patient timelines for the selection from B) showing low mean temporal toxicities. D) Patient projections with the toxicity layer. The selection from B) is highlighted with blue in this view, and shows moderate total severities for both \textit{acute} and \textit{late}. }
\label{fig:fig6}
\end{figure*}

\subsection{Case Study I: Multi-treatment Analysis}
The model builders wanted to find temporal symptom patterns across multiple treatment modalities and compare the results. The oncologists were hoping they would find specific symptoms highly correlated to specific treatment strategies. After examining the top row of the interface (\Cref{fig:teaser}.A), the evaluators noted that, unsurprisingly, common toxicities such as dryMouth, taste, swallow, and mucus were the highest overall (T 2.3). In general, symptoms usually followed a gradually increased toxicity during treatment and decreased post treatment, which was expected. However, symptoms related to daily life activities, such as mood, enjoyment of life, distress, and sadness showed severity peaks before the start of the treatment (i.e. first pink petal), implying that mental health improved when the patients started the treatment (i.e. the severity decreased). Next, the interface was used to show the symptom clusters for ICC (\Cref{fig:fig5}.A) and IRT (\Cref{fig:fig5}.B) in conjunction with the symptom queries (\Cref{fig:fig5}.C,D). Using the symptom queries, the evaluators found similar prevalent symptoms for both treatments (T2.3). In the symptom cluster components, both treatments showed two temporal clusters each, with similar overall symptom profiles (T2.1). Although the symptom queries showed many prevalent \textit{late} toxicities (\Cref{fig:fig5}.B,D), they were not all predicted by the model. These symptoms appeared as common \textit{late} toxicities in the rule mining results, as shown by the low opacity \textit{late} symptoms in the clusters panels (\Cref{fig:fig5}.A,B). Taste was predicted as a \textit{late} toxicity for ICC, correlated with the loss of appetite, and, surprisingly, with skin problems (\Cref{fig:fig5}.A). DryMouth showed obvious severe toxicity in \textit{late} when compared to the whole cohort (\Cref{fig:teaser}.A), more so for IRT (\Cref{fig:fig5}.B) (T2.1,3). The evaluators appreciated how the rose glyph projection kept symptoms with similar trajectories together. For instance, in the ICC symptom clusters, pain and mucositis showed strikingly similar trajectories (\Cref{fig:fig5}.A). They hypothesized this might be a sign that pain, being such a general symptom, was highly correlated with mucositis problems for this cohort. The evaluators also showed particular interest in the outliers of the \textit{acute} projections, namely problems with sleep in IRT and skin in ICC. 

Checking the Sankey diagrams for the two treatments (\Cref{fig:teaser}.C, \Cref{fig:fig7}), the evaluators observed that that IRT showed N3 stage (advanced) for node lymphs, while ICC did not present such a high attribute value (T2.4). Although the evaluated cohort had missing data, the oncology co-modelers appreciated the model’s ability to find common longitudinal patterns for small sub-cohorts which show increased risk of developing those patterns within the given treatment (T2.2). For example, although only 97 patients were given ICC (\Cref{fig:fig5}.A), the model predicted a higher likelihood (i.e. almost two times more likely) that appetite and skin problems could cause dryMouth as opposed to all the other treatments. The evaluators concluded that the symptom clustering component was an effective way to understand the impact of \textit{late} symptoms in a sub-cohort. They are excited to analyze the SRM results with more symptom rating data for this patient sub-cohort.

\subsection{Case Study II: Single Treatment Analysis}
For the second study, the oncology co-modelers wished to better understand the mechanisms between symptom clusters. They started with a treatment example, ICC. The interface was configured as follows: the symptom cluster component (\Cref{fig:fig6}.A), patient projection component using the symptom cluster layer (\Cref{fig:fig6}.B), the patient timeline component (\Cref{fig:fig6}.C), and the cohort characteristics component (\Cref{fig:teaser}.C). At first glance, the patients that usually suffered from moderate-to-high symptom burden overall showed patterns among the two existing symptom clusters (\Cref{fig:fig6}.A,B) (T1.1). Patients usually showed problems from both clusters, with lower burden patients sharing mostly cluster 1 ({appetite, skin} ->{taste}) (\Cref{fig:fig6}.B). Selecting the previously mentioned sub-group of patients with cluster 1 from the scatterplot (T1.4), the evaluators looked at their timelines (\Cref{fig:fig6}.C), and observed low mean symptom ratings for both treatment stages, with peaks among the symptoms from the clusters (T1.1). Moving to the cohort characteristics component, the modelers observed that most cluster combinations among this cohort showed higher symptom burden for the \textit{acute} and \textit{late} stages, but the symptom cluster 1 patients showed only below T3 stage problems (T1.3). While evaluating the cohort characteristics component, the oncology co-modelers commented that they expected severe symptom burden in \textit{late} to be correlated with higher T stage (i.e. T3) (\Cref{fig:fig6}.C), which this view proved that it was not the case.

Next, the evaluators wished to understand the overall temporal toxicity among patients better. The cohort characteristics component was changed to show the patient projection with the overall temporal severity layer applied (\Cref{fig:fig6}.D) (T1.2). This way, they could better understand the relationship between symptom cluster labels and \textit{acute}-\textit{late} toxicity. They noted that almost half of the patients within this treatment often showed severe toxicity during \textit{acute}, but low severity after the completion of treatment, which was received with relief. Selecting the top-left outlier (T1.4), the evaluators observed that the given patient did not have reported data for the \textit{acute} stage, making it an outlier, and agreed that the medical records for this patient needed further analysis. The oncology co-modelers expressed that the scatterplot was really efficient in detecting outliers in patient data, while also connecting the cohort to symptom burden characteristics. After finding the outliers and unexpected diagnostic patient details connected to symptom clusters, the evaluators decided that their future studies should focus further on diagnostic patient data.

\subsection{Expert Feedback}
The visual system and model results received positive feedback. One of the senior model builders affirmed: \textit{``The interface is extremely useful for navigating through the patient-reported outcome data and generating hypotheses. Evaluating the effect of different thresholds for symptom severity and rule mining would be overwhelming without these visualizations [...] Using the rose glyphs gives a quick overview of the symptom trajectory for a group of patients and it is easy to compare between different therapeutic combinations [...] The sequential rules provide a way to identify \textit{acute} symptoms that can be predictive for \textit{late} toxicity. The rule clustering dramatically reduces the complexity of the analysis by reducing the number of relevant rules and highlighting interesting metrics to compare the different treatments.”}

The oncology co-modelers were really impressed, one of them affirming: “The app is very good and combines all the information in one place, so that is very interesting”, while another commented: “I really like this..I feel very strongly about this, the utility for exploring the data here is very high” and “if you’re talking about quantitative decision- making, this is very strong”. The appreciation for multiple data-driven analyses was further emphasized by the oncologist co-modelers: “First, we can stratify by treatment, [...] second, we can see that patients who have certain patterns of symptoms like those more impacted by skin and appetite are more likely to get taste problems later on than [...] third, you can stratify the patients by T stage, N stage, and different clinical parameters [...] so for me, it is really, really helpful, it is a really cool tool”. \textcolor{black}{When considering wider adoption of the models in the clinic, the oncologists wished for additional workflows that started with data from the patient they are treating, and to analyze similar patients from the dataset to predict the \textit{late} symptoms for that patient. 
}

\textcolor{black}{Our visual encodings were designed through a parallel prototyping process, with frequent feedback and suggestions from collaborators, often aimed at reducing encoding complexity and increasing alignment with the clinical intuition (see Supplemental materials). In the end design,} the modelers appreciated the usefulness and many tasks that the rose glyphs accomplish, from single-symptom, single-treatment analysis to multi-symptom, multi-treatment analysis. One oncologist co-modeler commented when analyzing the rose glyphs: \textit{``Fascinating that taste is so prevalent [...] we don’t understand why it’s so bad. The kinetics are fascinating”}. \textcolor{black}{ The oncologists responded well to the inclusion of the rose glyphs as a fixed anchor at the top of the interface, and were able to immediately spot and comment on trends in different symptoms of interest. In an earlier iteration, a collaborator was able to identify a trend where patient symptoms decreased after the first week of treatment, which quickly led to finding an issue in the data preprocessing. Similarly, during the review, they immediately identified and commented on interesting trajectories for taste, and noted that they should explore taste-related issues in future studies (T2.1,T2.3). The glyphs’ horizontal separation for \textit{acute} and \textit{late} facilitated interpretation of sequential rule clusters (T1.1), and their compact design was particularly appreciated when comparing trajectories (T2.1, T2.3).} Secondly, they appreciated the Sankey diagram: \textit{``this one is going to help if you want to connect the dots between staging and toxicity, and symptom clusters, so it gives an overall connection”} \textcolor{black}{because they could compare symptom burden and clinical data across treatments more easily (T2.4). The diagram was an intuitive way of analyzing temporal and categorical patient attributes (T1.3) and it} revealed surprising results: \textit{``I expected that the more advanced staging you have (T stage), the more toxicity you get - it corrected my assumptions”}. They found the scatterplot useful to observe symptom burden temporal trends at the cohort level while detecting outliers\textcolor{black}{, and to compare symptom burden trends across treatments (T1.4, T2.3). As our collaborators routinely analyze cohorts using scatterplots, they found the scatterplot temporal abstraction intuitive to analyze overall symptom burden and its relation to symptom clusters (T2.2). } The other components served as useful complements to the model analysis. 

\section{Discussion}
This work was developed as a collaborative project alongside oncologists and data scientists to create explainable rule mining and clustering of temporal patient symptoms. The evaluation with domain experts in symptom research demonstrates that our visual system successfully explains the SRM model results in the context of several aspects of the patient and symptom cohort data. Our results show that our visual system is an effective tool for collaboratively analyzing treatment-related symptom patterns in clinical patients. Our combination of SRM and rule clusters allows for a comprehensible explanation of common co-occurring symptoms and predicting \textit{late} stage symptoms for different treatment groups. While we focus our design on model building, our case studies and feedback suggest that our interface is able to provide usable insights for clinical practitioners. Although we target radiation oncology patients, we generalize design insights to a wide range of approaches when dealing with complex, temporal patient outcomes and when working with clinical explainable ML models. 
Next, we present the lessons learned from this multi-disciplinary collaboration:

\textbf{L1.} \textit{Use visual scaffolding~\cite{marai2015visual} to introduce new visual encodings.} At the beginning of the design process, we started by visually listing sequential rules to the senior modelers, which were hard to interpret due to too many existing patterns. Thus we worked on a model that would summarize the mined rules, but we needed the means to convey the rule results in a meaningful way. That is how we came up with the rose encoding, which in the beginning, provoked some skepticism from the domain experts \textcolor{black}{who were used to the rule abstraction from our previous work, which used node-link encodings for rules, their antecedents, and consequents. However, this abstraction did not work in the present work because node-link representations were not able to deal with large numbers of rules or capture temporal patterns. Replacing the original node-link with 2D projections, whose items were temporally separated and grouped using envelopes, proved to be an intuitive solution.} After a couple of sessions throughout the interface prototyping stage, the oncology co-modelers got to rely on this encoding the most, and during the evaluation session, it ranked the highest among their preference. 

\textbf{L2.} \textit{Focus on actionability and transparency when working with clinical XAI applications.} When developing our model, we focused on rule-mining based on the positive reception the approach received from clinical researchers due to its simplicity and transparency. However, we found that large rule sets with overlapping results made the model lack actionability. We addressed this issue by producing rule clusters that could be easily interpreted. \textcolor{black}{Moreover, adapting rule metrics (i.e. support, confidence, lift) to clinical context (i.e. symptom risk) helped the team identify interesting results, adding to the actionability of the SRM models. } This drastically improved reception from collaborators.

\textbf{L3.} \textit{Use highly configurable interfaces in XAI modeling.} \textcolor{black}{ Although our designed focused on visually interpreting the SRM clusters and evaluate how they are impacted by treatments, we found that properly analyzing the data required varied workflows and orientations, such as analyzing individual patients, rules, symptoms, treatments, and clusters, in order to fully understand the underlying algorithm and assess where issues may arise. } Our human-machine visual system supports a variety of workflows with the help of six visual components. \textcolor{black}{At the same time, the use of configurable layouts allowed us to minimize cognitive load when working collaboratively by hiding unnecessary components.}

\textbf{L4.} \textit{Account for multiple levels of details when working with collaborative workflows.} Our system was designed in coordination with multiple domain experts, who approach the problem with different viewpoints, \textcolor{black}{which required different forms of data abstraction}. For example, a senior model builder was more interested in identifying the rules with the highest confidence and support, \textcolor{black}{and thus benefited from views with higher levels of aggregation such as the symptom query view, along with a layout that was more focused on showing multiple different panels. In contrast, the} oncologists gave insights into potential mechanisms behind symptom clusters, and others were interested in exploring single patients to identify and explain outliers or assess the value of the rules when explaining results to patients\textcolor{black}{, and thus benefited more from the inclusion of the scatterplots alongside the symptom cluster view. In addition, since clinicians were more interested in the impact of different treatment groups, they benefited more from configuring the layout to allow for side-by-side comparisons between panels showing results for different treatment groups.} By providing a configurable interface that allowed for analysis of both rule sets and sub-groups of patients with different granularities, we were able to better accommodate different insights \textcolor{black}{and workflows }from experts in data mining and oncology.

Because our system aims to visualize individual patients in the cohort, some of our visual components such as the scatterplot and individual patient timelines can show scalability issues if they must support a large number of patients (e.g. n > 800). However, this may be addressed by increasing the granularity of the sub-cohorts used to reason about the data. On the other hand, the Sankey diagram, rose glyphs, and symptom query barcharts can support any cohort sizes. \textcolor{black}{The timeline component aims to support analysis of only one or a handful of patients at a time. }Moreover, if having to support more data attributes, the Sankey diagram would become harder to understand, although brushing operations can uncover the necessary connections. On the other hand, given the difficulty in collecting large homogenous cohorts of symptom data, we felt that it was more important to provide a highly configurable interface, supporting several workflows, at the cost of some scalability issues. \textcolor{black}{ Each visual component and view of the interface can be initialized with a given sub-cohort, with consistent layouts across the views, in order to support side-by-side symptom cluster, symptom burden, or treatment comparison. Single-cohort, symptom, treatment, and outlier analysis is further supported across views through brushing and linking operations. Pairwise sub-cohort comparison was, however, not a modeling goal. 
}

Notably, some of the patients used in the model building were still under the observation period, and as a result, they were missing symptom ratings for many after treatment time points. This impacted the results of the model’s predictions. \textcolor{black}{Our modeling approach is generalizable, although it is a clinical-practice based model. Future work includes supporting SRM model refinement once surveillance is complete, and applying rule mining to longitudinal treatment plans even after potential cancer recurrence.}

\section{Conclusion}
In this work, we described the activity-centered design of a visual analytics system that helps to explain and validate our proposed multivariate model for longitudinal symptom analysis. While we examine a cohort of HNC patients, our approach can be generalized to other disease applications that study cohort toxicities. Our back-end uses SRM in conjunction with other unsupervised methods to predict and find temporal patterns in cancer symptoms, while our front-end supports the analysis of these models in the context of real patient data. We propose SRM to find temporal symptom clusters and a new visual encoding, the rose glyphs, to describe the resulting clusters and predictions. Our visual system supports various workflows through configurable components, which guide to a better understanding of treatment-related symptoms for multiple treatments. The evaluation with domain experts in cancer symptom model building demonstrates the usefulness of our approach in clinical research. Lastly, we summarize the lessons learned from this multidisciplinary collaboration, and we hope they will guide towards better XAI applications in healthcare.


\acknowledgments{%
	Our work is supported by NIH awards
NCI-R01-CA258827 and NLM-R01-LM012527, and NSF awards
CDSE-1854815 and CNS-1828265.%
}

\bibliographystyle{abbrv-doi-hyperref}

\bibliography{template}

\end{document}